\documentclass[twocolumn,final,aps,prl,showpacs,superscriptaddress,amsmath,amssymb,amsfonts,floatfix]{revtex4-1}

\usepackage{graphicx}
\usepackage{multirow}
\usepackage{epsfig}
\usepackage{url}
\usepackage{color}

\usepackage[colorlinks,breaklinks,bookmarks=true,citecolor=blue,linkcolor=blue,urlcolor=blue]{hyperref}
\usepackage{color}
\usepackage{tabularx}
\usepackage{sidecap}
\usepackage{soul}
\usepackage{float}

\usepackage{orcidlink}

\begin{document}

\author{Liang Si\,\orcidlink{0000-0003-4709-6882}}
\email{liang.si@ifp.tuwien.ac.at}
\affiliation{School of Physics, Northwest University, Xi'an 710127, China}
\affiliation{Institute of Solid State Physics, TU Wien, 1040 Vienna, Austria}

\author{Xiaochao Wang}
\affiliation{School of Physics, Northwest University, Xi'an 710127, China}

\author{Paul Worm\,\orcidlink{0000-0003-2575-5058}}
\affiliation{Institute of Solid State Physics, TU Wien, 1040 Vienna, Austria}

\author{Wei Peng}
\affiliation{Department of Physics, University of Warwick, Coventry CV4 7AL, U.K.}

\author{Minjae Kim}
\affiliation{Korea Institute for Advanced Study, Seoul 02455, Republic of Korea}

\author{Lingfei Wang}
\affiliation{Hefei National Research Center for Physical Sciences at Microscale, University of Science and Technology of China, Hefei, 230026 China}

\author{Karsten Held\,\orcidlink{0000-0001-5984-8549}}
\affiliation{Institute of Solid State Physics, TU Wien, 1040 Vienna, Austria}

\title{Chiral magnetism and ordering of oxygen vacancies  in SrTiO$_{2.5}$}

\begin{abstract}
Oxygen vacancies in the perovskite insulator 
SrTiO$_3$ free electrons that couple with other physical degrees of freedom such as lattice, orbital, and spin. This leads to the emergence of exotic quantum states such as superconductivity and unusual ferromagnetism. We perform density-functional theory and dynamical mean-field theory calculations and demonstrate that the orientation and ordering  of the TiO$_5$ pentahedra  plays a crucial role. Specifically, for vacancy-rich   SrTiO$_{3-\delta}$ ($\delta\sim$0.5), we find a chiral ordering of the TiO$_5$ pentahedra in a  sixfold superlattice. This  chiral structure is  accompanied by a chiral magnetic state with a net moment in the (111) direction at room temperature, which can explain  several experimental observations.  
\end{abstract}

\maketitle
Perovskite transition-metal oxides with a pseudocubic structure are ubiquitous and have been studied intensively. The coupling between distinct physical degrees of freedom, such as spin, charge, and orbital, together with correlations, heterostructuring and atomic-level defects \cite{hwang2012emergent,tokura2000orbital,georges2013strong} leads to exotic quantum states, including colossal magnetoresistance \cite{Kusters1989,vonHelmholt1993,ramirez1997colossal,Millis1995,Held2000}, topological skyrmions \cite{matsuno2016interface,wang2018ferroelectrically,lu2021defect}, high-mobility two-dimensional electron gas \cite{ohtomo2004high,PhysRevLett.98.216803}, unusual ferromagnetism \cite{lee2013titanium,bi2014room}, electronic phase separation \cite{ariando2011electronic}, and interface superconductivity \cite{reyren2007superconducting,li2011coexistence}. The physical properties of perovskite oxides can be altered by introducing various defects, such as vacancies or atomic substitution. For example, oxygen vacancies (V$_{\mathrm O}$) can induce electron doping; and their amount can be controlled by annealing samples at reduced oxygen pressure, thus synthesizing SrTiO$_{3-\delta}$ (STO$_{3-\delta}$) \cite{rao2014laser}.

As a consequence of its unique properties, a chemically stable  band insulator and centrosymmetric paraelectric \cite{van2001bulk}, STO$_3$ is {\em the} generic substrate for growing perovskite-structure $AB$O$_3$  films or heterostructures \cite{PhysRevB.75.121404,wang2012interface,ariando2011electronic,li2011coexistence,lee2013titanium,ohtomo2004high,li2019superconductivity}.
Effects of V$_{\mathrm O}$ in STO$_{3-\delta}$ have been intensively studied \cite{PhysRevB.47.8917,tan2014oxygen,astala2001ab,jiang2011mobility,wang2013oxygen,muller2004atomic,choi2013anti,hanzig2016anisotropy,alexandrov2009first,Behrmann2015,sing2017influence,schie2012molecular,jeschke2015localized,hou2010defect,chan1981nonstoichiometry,eom2017oxygen,PhysRevLett.111.217601,PhysRevLett.98.115503}. 
Here, experimental \cite{PhysRevLett.107.256601} and theoretical research \cite{PhysRevLett.111.217601,lopez2015research,hou2010defect,PhysRevB.92.115112,PhysRevB.85.020407,PhysRevB.86.064431} revealed signatures of both localized and delocalized physics: the former is due to the tendency
of one of the two electrons, which are released by the missing O$^{2-}$, to localize in a bonding state formed between the two Ti sites adjacent to the V$_{\mathrm O}$; while the latter is caused by the second electron occupying a delocalized state composed of Ti-$t_{2g}$ conduction bands \cite{PhysRevB.85.020407,PhysRevB.86.064431}.
When a low carrier density of the order of 10$^{17}$\,cm$^{-3}$ is induced by V$_{\mathrm O}$, the insulator STO$_3$ turns into an $n$-type conductor \cite{gor2016phonon} and even a superconductor at higher doping levels and lower temperatures ($T<$1\,K) \cite{PhysRevLett.12.474,PhysRev.163.380,gastiasoro2020superconductivity,PhysRev.106.162}.
Another well-studied  phenomenon is anomalous ferromagnetism in STO$_{3-\delta}$: signals of ferromagnetism in the bulk, on the surface and at interfaces  have been observed in measurements of the anomalous Hall effect \cite{PhysRevLett.106.136809}, Kondo scattering \cite{PhysRevLett.109.196803,PhysRevLett.107.256601}, polarization dependent x-ray absorption spectroscopy (XAS) \cite{PhysRevLett.111.087204} and SQUID \cite{rao2014laser} both at high \cite{ariando2011electronic,PhysRevB.87.220405,trabelsi2016effect,rao2014laser,bi2014room} and low temperatures \cite{PhysRevLett.111.087204}.

The observed ferromagnetism in STO$_{3-\delta}$ has been attributed to the existence of V$_{\mathrm O}$ near the interface \cite{brinkman2007magnetic} and surface \cite{coey2016surface} based on the following observations: (1) Magnetic moments in  STO$_{3-\delta}$ are quenched upon annealing in an oxygen-rich atmosphere, indicating the indispensable role of V$_{\mathrm O}$ in the formation of ferromagnetism \cite{rao2014laser}. (2) In Nb-doped STO$_3$, ferromagnetism can be eliminated (recovered) by annealing in air (vacuum) \cite{PhysRevB.87.220405}. This indicates again that  charge carrier doping is related to the emergent ferromagnetism and V$_{\mathrm O}$ are essential. (3) The clustering of V$_{\mathrm O}$ in STO$_{3-\delta}$ gives rise to the formation of midgap state(s), that are beyond a rigid-band approximation \cite{PhysRevB.57.2153}. These midgap states have the character of Ti-3$d_{z^2}$ whose energy becomes lower (due to the formation of a bonding state) than the other Ti-3$d$ conduction bands and are thus preferably occupied by the released electrons that form a local magnetic moment. 
However, a complete physical picture connecting the formation of room-temperature ferromagnetism and V$_{\mathrm O}$ in STO is still missing.

\begin{figure}[t]
\centering
\includegraphics[width=\columnwidth]{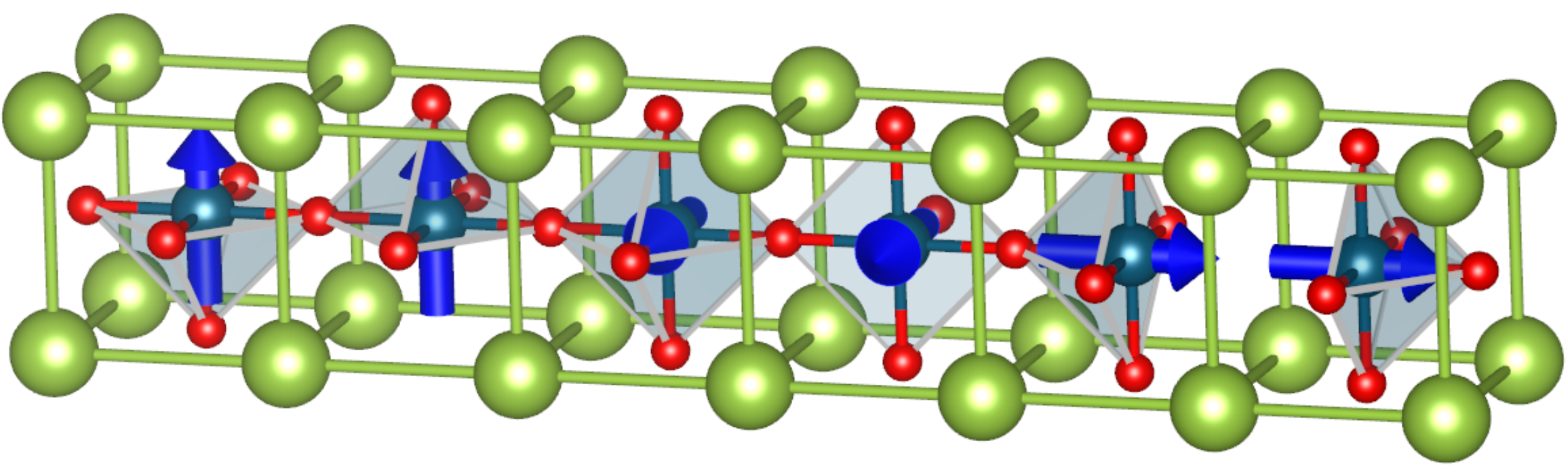}
\caption{Schematic figure of chiral order of oxygen vacancies and Ti magnetic moments in SrTiO$_{2.5}$ (schematics). The latter have a right-handed orientation and a net ferromagentic moment along the diagonal (cubic)  direction 
}
\label{fig:schematics}
\end{figure}

In this paper, we consider possible relevant crystal structures of STO$_{3-\delta}$ at a defect density $\delta\sim0.5$ \footnote{According to the previous work \cite{zhu1997effects,chehrouri2018novel}, for STO$_{3-\delta}$ at $\delta\sim$1 the structure hosts infinite-layer structure of SrTiO$_2$. Hence we limit all the possible V$_{\mathrm O}$ positions to the out-of-plane ones (apical position above/under Ti ions), i.e., the positions created by removing O from SrO layers.}. 
We find a long periodic arrangement of TiO$_5$ pentahedra in a chiral order of oxygen vacancies (COV), see Fig.~\ref{fig:schematics}. 
This structural spiral  entails a magnetic spiral, see again Fig.~\ref{fig:schematics}, where every two magnetic moments from left to right are arranged in a right-hand fashion. It has a ferromagnetic net moment in the (111) direction, which, thanks to strong correlations, is stable  even at room temperature.


\begin{figure*}[t]
\centering
\includegraphics[width=\textwidth]{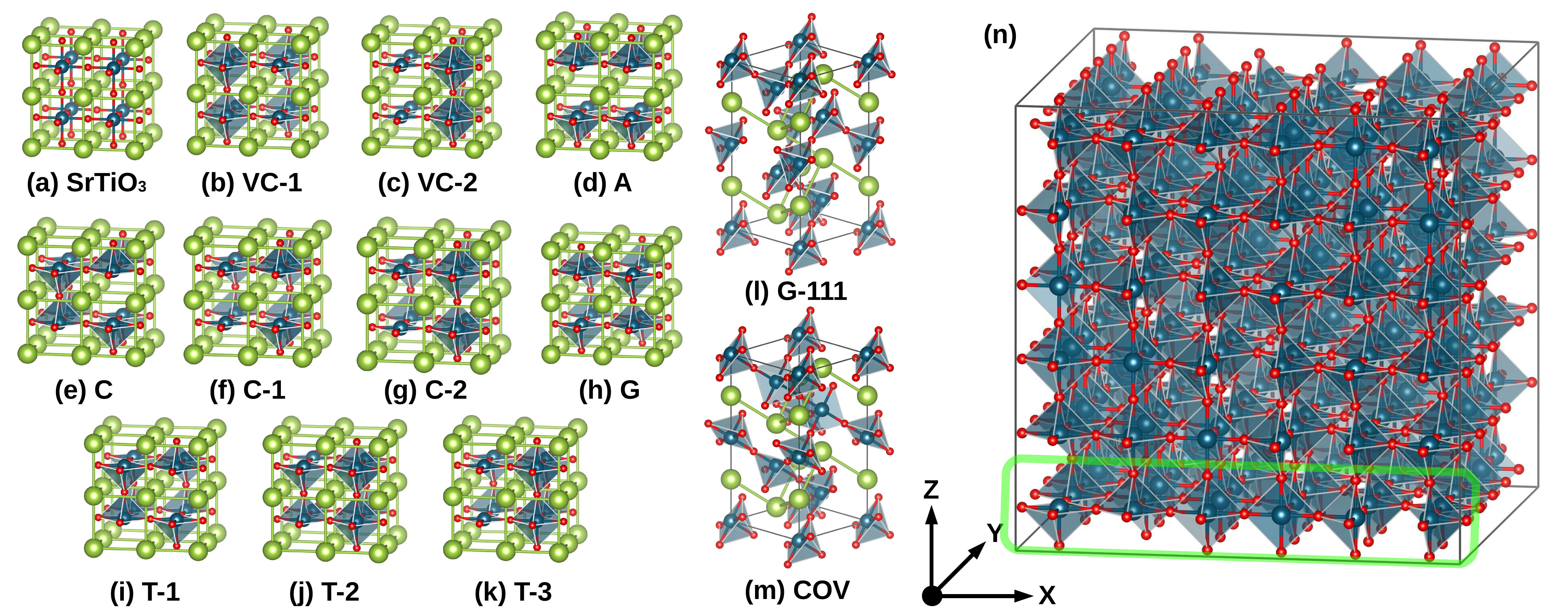}
\caption{Possible crystal structures of STO$_{2.5}$ (a-m). The cubic coordination of the COV (m) phase is shown in (n), requiring a 6$\times$6$\times$6 supercell. The green line indicates that in $x$/$y$/$z$ direction the TiO$_5$ pentahedron are ordered in a long range  as: +$Z$, -$Z$, +$Y$, -$Y$, +$X$, -$X$. Please note that the $G$-phase (h) is structurally equivalent to the $G$-111 phase (l), the latter is an effective representation of the former in hexagonal coordination ($a$=$b$$\neq$$c$, $\alpha$=$\beta$=90$^\circ$, $\gamma$=120$^\circ$). Please note the additional rotations of the pentahedra in the COV phase (m) compared to the $G$-111 phase (l).}
\label{Fig1_structure}
\end{figure*}

\paragraph{Oxygen vacancy structure.}
The possible structures of STO$_{2.5}$ within a 2$\times$2$\times$2 supercell or hexagonal cells are shown in Fig.~\ref{Fig1_structure}(b-k) with Fig.~\ref{Fig1_structure}(a) showing the defect-free  supercell. 
To achieve STO$_{2.5}$, four V$_{\mathrm O}$ need to be accommodated in the supercell (we here only consider V$_{\mathrm O}$'s in the SrO$_2$ layers as this is energetically favorable; for computational details of DFT and DMFT calculations see Supplementary Materials (SM) \cite{SM} Section~I-IV).
Fig.~\ref{Fig1_structure} (l,m) show two additional 1$\times$1$\times$6 hexagonal supercells in the cubic (111) direction with 3 V$_{\mathrm O}$'s.
Previously, theoretical \cite{PhysRevLett.98.115503} and experimental \cite{muller2004atomic} studies reported a clustering tendency of V$_{\mathrm O}$ in STO$_{3-\delta}$: i.e.,  an ``apical divacancy'' cluster along the $z$-direction is more stable than other configurations in DFT(+$U$)  \cite{PhysRevLett.98.115503}. 
The structures in Fig.~\ref{Fig1_structure}(b,c) [named as vertical chain-1 and -2 ($VC$-1 and $VC$-2), respectively] correspond to  such a clustering tendency.
The $A$ structure [Fig.~\ref{Fig1_structure}(d)] is obtained by removing all the oxygen from a selected SrO layer in the supercell.
In Fig.~\ref{Fig1_structure}(e,h), the $C$ and $G$ structures are obtained by orienting the TiO$_5$ pentahedrons differently, while the coexistence of TiO$_4$ planer and TiO$_6$ octahedron is forbidden.
For the $C$-1, $C$-2, $T$-1, $T$-2 and $T$-3 structures, coexistence of TiO$_4$ planar, TiO$_5$ pentahedron and TiO$_6$ octahedron is allowed.
In the $G$ structure [Fig.~\ref{Fig1_structure}(h)], the V$_{\mathrm O}$-Ti-O direction has, in fact, the same form as spin-up and spin-down in a $G$-type antiferromagnetic order. This allows a transformation from the $G$ to the $G$-111 phase [Fig.~\ref{Fig1_structure}(l)]: $G$-111 is hexagonal 
in the cubic (111) direction,
while $G$ is exactly the same structure as $G$-111 (rotated and  in a larger supercell with 8 instead of 6 Ti atoms). Based on $G$-111 we additionally obtain one more structure: the COV phase [Fig.~\ref{Fig1_structure}(m)] that is obtained by rotating the V$_{\mathrm O}$-Ti-O directions in Fig.~\ref{Fig1_structure}(l) by 120$^{\circ}$ along the (001) direction of the 1$\times$1$\times$6 hexagonal primitive cell, in a chiral way from layer to layer.

\begin{figure}[t]
\centering
\includegraphics[width=0.925\columnwidth]{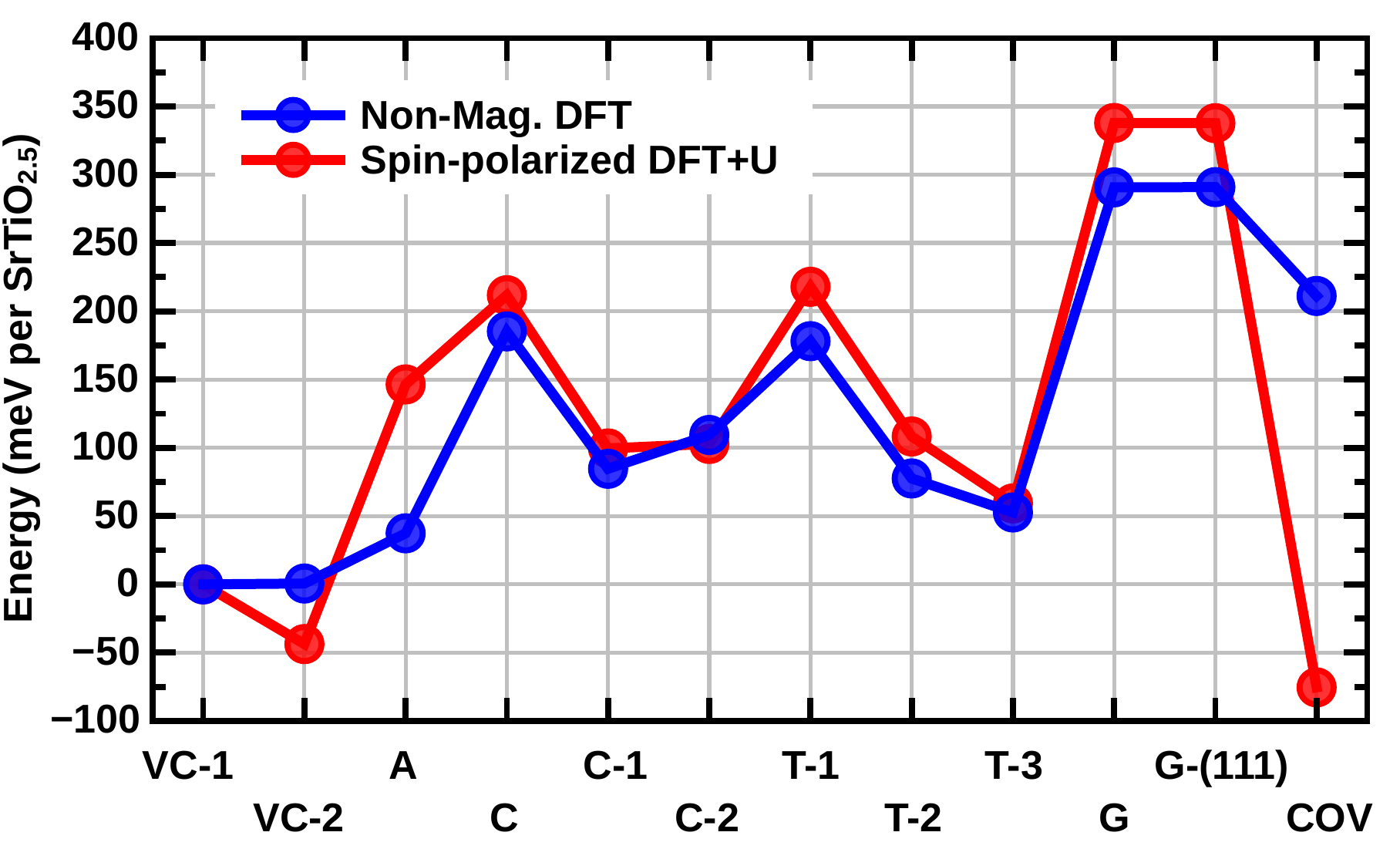}
\caption{(a) DFT (PBESol)  total energies for the different phases of Fig~\ref{Fig1_structure}. For both non-spin-polarized DFT and spin-polarized DFT+$U$ calculations, the energies of the $VC$-1 phase are set to zero as a reference.}
\label{Fig2_energy}
\end{figure}

Let us now turn to the analysis of the structural stability using DFT(+$U$). In both non-spin-polarized DFT and spin-polarized DFT+$U$ calculations, $VC$-1 and $VC$-2 with both planar TiO$_4$ and octahedral TiO$_6$ layers are stable compared to all other 2$\times$2$\times$2 structures [Fig.~\ref{Fig1_structure}(a-k)], see   Fig.~\ref{Fig2_energy}  and Table~S.I in SM Section~II. 
However, there is another phase that has an even lower energy in DFT+$U$: the COV phase becomes the ground state being -75.42\,meV/STO$_{2.5}$ lower in energy than the $VC$-1 phase, which is the ground state in DFT.
It is also -31.71\,meV/STO$_{2.5}$ lower in energy than  $VC$-2 which is 
preferable to $VC$-1  in DFT+$U$.

From Fig~\ref{Fig1_structure}(l,m) one can see that the only difference between the $G$-(111) and the COV structure is the orientation of the TiO$_5$ pentahedrons:
in the $G$-(111) structure all the pentahedra point in the (001) cubic direction  [Fig.~\ref{Fig1_structure}(h)], i.e., the (112) hexagonal direction [Fig.~\ref{Fig1_structure}(l)].
In contrast, in the COV phase, the TiO$_5$ pentahedrons order in a more complicated chiral way, and in a chiral space group of $P3_221$ (154). To have a better understanding of this long-range structural periodicity, the COV phase is shown in its original cubic coordination in Fig~\ref{Fig1_structure}(n). As shown in the region marked by a green rectangle in Fig~\ref{Fig1_structure}(n) (which is also in Fig.~\ref{fig:schematics}), the TiO$_5$ pentahedra alternate their orientations in all possible directions with a pattern +$Z$, -$Z$, +$Y$, -$Y$, +$X$, -$X$.
That is, the even pentahedra in this chain are oriented left-handed and the odd ones a right-handed. 
For accommodating this structure in a cubic structure, a 6$\times$6$\times$6 supercell of Fig~\ref{Fig1_structure}(n) that contains 972 atoms is required. 

The relaxed primitive cell of the COV phase STO$_{2.5}$ in Fig.~\ref{Fig1_structure}(m) is hexagonal with a pseudocubic lattice constant of 3.909\,\AA~or a $c$-lattice constant 13.53\,\AA~in the cubic coordination of Fig.~\ref{Fig1_structure}(m). This is consistent with experimental reports on STO$_{2.5}$: (1) X-ray diffraction (XRD) pattern revealed STO$_{2.5}$ could be interpreted as from a cubic cell with a lattice constant of 3.903\,\AA~\cite{franco1977anion}; (2) a sixfold superlattice along (111)$_{cubic}$ orientation was observed, suggesting a long periodic spacing of 13.52\,\AA~\cite{tofield1978anion}. This is consistent with our DFT relaxed out-of-plane $c$-lattice of the COV phase [Fig.~\ref{Fig1_structure}(m)]: 13.53\,\AA. To further study the dynamical stability of COV STO$_{2.5}$, we compute its phonon spectra and the corresponding phonon DOS shown in Fig.~S.1 in SM Section~II.

\paragraph{Magnetic structure.}
The reason why the COV structure is energetically the most favorable in DFT+$U$ is the formation  of a large magnetic moment of 1.0\,$\mu_B$/Ti, i.e., the released electrons are fully spin-polarized [note STO$_{2.5}$ has Ti$^{3+}$ cations (3$d^1$ state) on all lattice sites]. This is opposite to the conclusions from previous research reporting the coexistence of localized and delocalized electrons in STO$_{3-\delta}$ for smaller $\delta$  \cite{PhysRevLett.107.256601,PhysRevLett.111.217601,lopez2015research,hou2010defect,PhysRevB.92.115112}.


\begin{figure*}[t]
\begin{minipage}{.795\textwidth}
\centering
\includegraphics[width=0.95\textwidth]{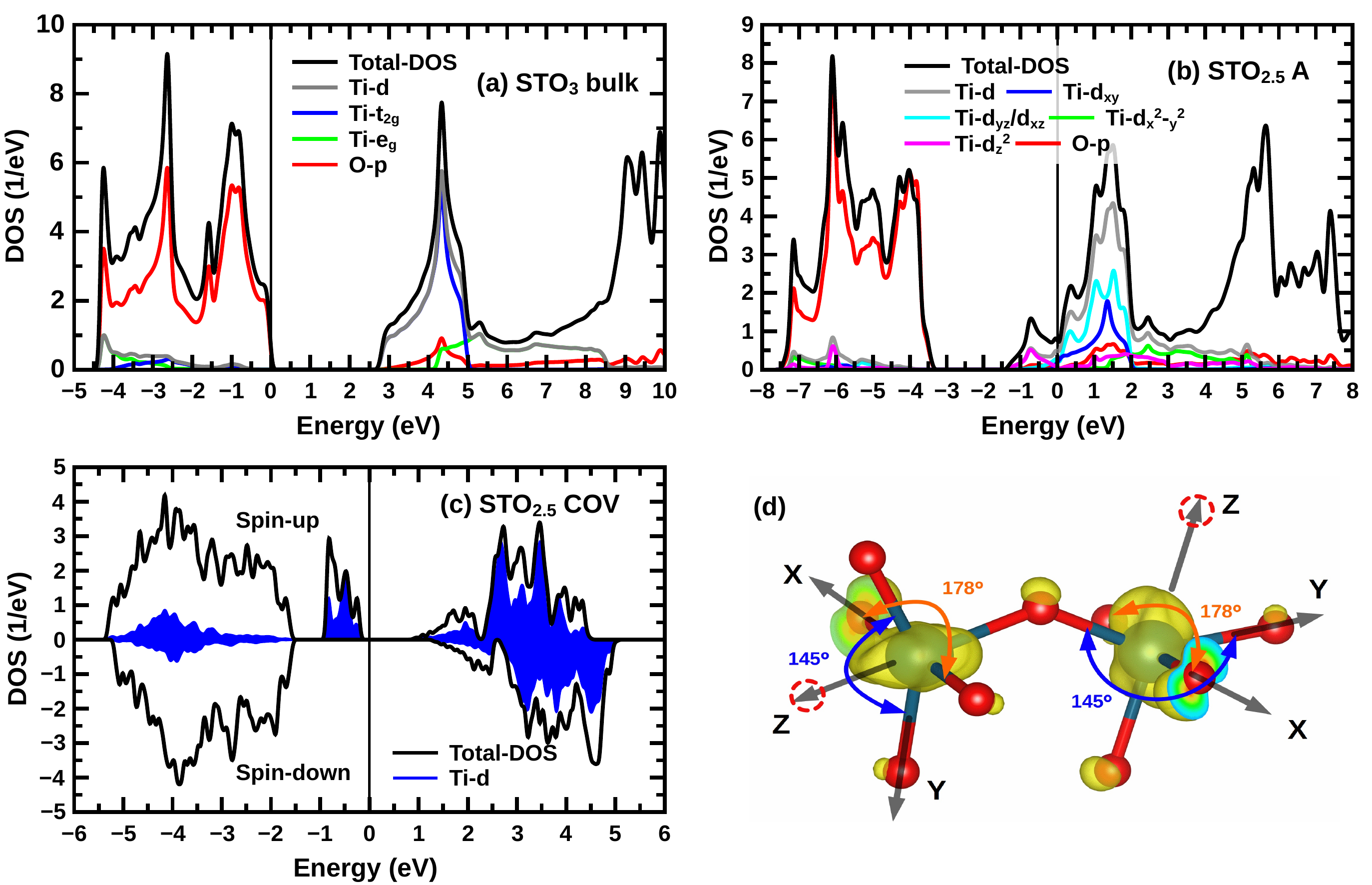}
\end{minipage}\hfill
\begin{minipage}{.195\textwidth}
\caption{DFT(+$U$) DOS of (a) bulk STO$_3$  [Fig.1~\ref{Fig1_structure}(a)] with lattice constant $a$=3.905\,\AA, (b) STO$_{2.5}$  $A$ phase [Fig.~\ref{Fig1_structure}(d)], (c)  STO$_{2.5}$  COV phase  [Fig.~\ref{Fig1_structure}(m)]. (d) Band-resolved partial charge densities of the occupied states in the spin-majority channel of the  STO$_{2.5}$ COV phase. Here, the isosurface is set as 0.004 and in units of e/\AA$^2$, the dashed circles in (d) indicate the position of the V$_{\mathrm O}$; for each TiO$_5$ pentahedron a local coordinate system is identified. 
}
\label{Fig3_STO_DFT}
\end{minipage}
\end{figure*}

Fig.~\ref{Fig3_STO_DFT}(a) displays the electronic DOS of bulk STO$_3$ using the mBJ functional \cite{PhysRevLett.102.226401}, which predicts a band gap of $\sim$2.6\,eV. While smaller than the 3.2\,eV reported in experiments \cite{van2001bulk}, standard DFT-PBE would predict a band gap of only 1.88\,eV. Next we study the simplest case of V$_{\mathrm O}$ in STO$_{2.5}$: the $A$ phase [Fig.~\ref{Fig1_structure}(d)]. Here, the oxygen atoms are removed from a selected SrO layer, which also corresponds to a single V$_{\mathrm O}$ in a STO 1$\times$1$\times$2 supercell. The structure is predicted to be non-magnetic. The overlap between the two $d_{z^2}$ orbitals of Ti (adjacent to V$_{\mathrm O}$) results in bonding and anti-bonding states. One of the two additional electrons is localized in the $d_{z^2}$ bonding state (see the DOS of $d_{z^2}$ from -1\,eV to 0\,eV).  The second electron is delocalized and occupies the $t_{2g}$ conduction bands that now cross the Fermi energy  E$_f\equiv 0$ [see the DOS of $d_{yz}/d_{xz}$ and $d_{xy}$ between -0.5\,eV to 0\,eV in Fig.~\ref{Fig1_structure}(b)] \cite{PhysRevLett.111.217601}. 
This is precisely the previous scenario of coexistence of localized and delocalized states.

Let us now turn to the DFT+$U$ spectrum of the STO$_{2.5}$ COV phase. A gap opens at E$_f$ and the full occupation of the spin-majority (up) channel shows ferromagnetism with a magnetic moment $\sim$1.0\,$\mu_B$/Ti [Fig.~\ref{Fig3_STO_DFT}(c)].
The magnetic moments of all other phases are considerably smaller  (see Table~S.I in SM Section~III). For example, the magnetic moments for the $VC$-1  and $VC$-2 phases
are 0.3 and 0.4\,$\mu_B$/Ti, respectively, consistent with previous report \cite{zhang2012vacancy} of $\sim$0.42\,$\mu_B$/Ti. 
In the COV phase, the six released electrons of the 1$\times$1$\times$6 primitive cell are trapped in six in-gap bands, corresponding to the DOS between -1\,eV to 0\,eV in Fig.~\ref{Fig3_STO_DFT}(c). These states are well separated from the unoccupied Ti-$d$ bands in the spin-majority channel. 
These fully localized states from 0 to -1\,eV predicted by DFT+$U$ for STO$_{2.5}$ indicates that the conductivity of STO$_{3-\delta}$ may decrease  when $\delta$ goes from 0 to $\sim$0.5, \emph{cf.}, experiments for smaller $\delta$ \cite{gong1991oxygen}.

Another noteworthy observation from  Fig.~\ref{Fig3_STO_DFT}(c) is that the gap of the V$_{\mathrm O}$ phase is only from 0 to $\sim$1\,eV to the first unoccupied states with the maximum of the unoccupied states at 2.2\,eV. This boosts an emission around 2 to 3\,eV, being consistent the blue-light emission at $\sim$2.8\,eV observed in STO$_{3-\delta}$ layers \cite{kan2005blue}.
Please note that DFT+$U$ may lead to an overestimation of the band gap for (anti-)ferromagnetic insulators; hence smaller gaps are expected in realistic samples or computations with $GW$+DMFT or DFT+DMFT \cite{held2011hedin}. We further compute the DFT+$U$ total energy of the antiferromagnetic (AFM) order. Setting the COV phase as inter-layer AFM state, we find that the FM coupling is more stable than AFM order by $\sim$10\,meV per STO$_{2.5}$ formula, indicating FM order is the ground state. After switching on spin-orbital coupling (SOC) and non-colinear magnetic order, the chiral order shown in Fig.~\ref{fig:schematics} is slightly modified to the out-of-plane (111) direction, see the DFT+$U$+SOC results in Fig.~S2 in SM Section~IV).

Next, we analyze the orbital characters of the occupied bands in Fig.~\ref{Fig3_STO_DFT}(c) by plotting the band-resolved partial charge densities of the occupied states for the two neighboring Ti cations in Fig.~\ref{Fig3_STO_DFT}(d). Only the spin-majority states are plotted and we use an energy window of -1 to 0\,eV. Here, local coordinates define the TiO$_4$ plane as the $xy$-plane while the V$_{\mathrm O}$-Ti-O vector serves as the $z$-axis. As mentioned, the occupied states have a $d_{xz}$ orbital character, indicating orbital ordering. 

\begin{figure}[t]
\centering
\includegraphics[width=\columnwidth]{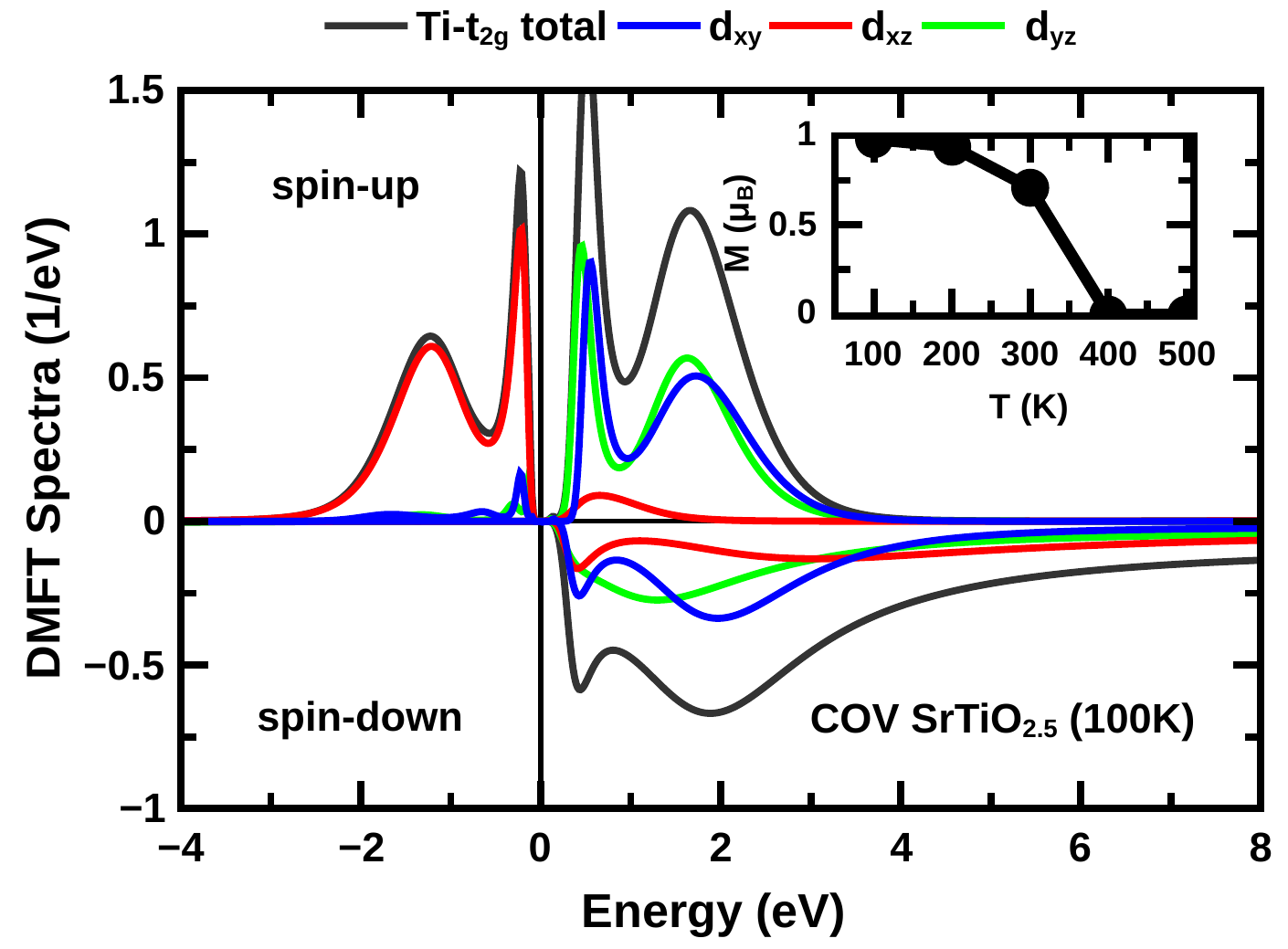}
\caption{DMFT spectral functions $A(\omega)$ of the COV phase STO$_{2.5}$ at 100\,K. Inset: magnetic moments as a function of temperature.}
\label{Fig4_STO_DMFT}
\end{figure}

\paragraph{Dynamical electronic correlations.}
DFT(+$U$) is reliable for ground state calculations in the presence of static magnetic (or orbital) order.
In the following, we include dynamic correlations using DFT+DMFT.
The DMFT spectral functions $A(\omega)$ of COV STO$_{2.5}$ at 100\,K are shown in Fig.~\ref{Fig4_STO_DMFT}, showing a FM semiconductor with a gap of $\sim$200\,meV.
Similar to DFT+$U$, the occupied spectrum only contains the spin-majority channel of the $d_{xz}$ orbital [for detailed discussion and momentum-resolved DMFT spectra $A(k,\omega)$ see SM Section~V and Fig.~S3].

In an idealized pentahedron crystal environment, the five $d$-bands split into four groups (from lower to higher energies): $d_{yz}$/$d_{xz}$, $d_{xy}$, $d_{z^2}$ and finally $d_{x^2-y^2}$. The released electrons from V$_{\mathrm O}$ are expected to occupy the double degenerate $d_{yz}$/$d_{xz}$ orbitals in an undistorted pentahedron. However, in the COV phase, the pentahedra are distorted because of its long-range period [for the details of relaxed structures see Fig.~\ref{Fig3_STO_DFT}(d)]. Within the TiO$_4$ ($xy$) plane, the O-Ti-O bond (the one along $y$) shrinks to 145$^{\circ}$ while another one is basically unchanged (178$^{\circ}$; along the $x$-axis). This bond bending compresses the $d_{yz}$ orbital and increases its energy. As a result, the degeneracy of $d_{yz}$/$d_{xz}$ is lifted; an orbital-ordered state with occupied $d_{xz}$ orbitals forms, see Fig.~\ref{Fig4_STO_DMFT}. Because of the change of the local coordination axis [see Fig.~\ref{Fig3_STO_DFT}(d)] this is akin to an alternating orbital order, which leads to ferromagnetism because hopping in a super-exchange process to a local spin-1 is favored by Hund's exchange \footnote{Indeed the interplay of ferromagnetism and orbital ordering has been studied since the dawn of DMFT~\cite{Held1998}.}. The FM spin-splitting further lowers the energy of the occupied parts and leads to a magnetic moment of 1.0\,$\mu_B$/Ti. 

To investigate the temperature dependence of the FM phase, we compute the magnetic moment of COV STO$_{2.5}$ vs.\  temperatures and plot it in the inset of Fig.~\ref{Fig4_STO_DMFT}. The moment at 100\,K is $\sim$1.0\,$\mu_B$/Ti, and gradually decreases to 0.93\,$\mu_B$ and 0.78\,$\mu_B$ at 200\,K and 300\,K, respectively. Finally, it vanishes at 400\,K, indicating that the transition to the paramagnetic state is above room temperature.

\paragraph{Conclusion.} 
We found  a COV phase in STO$_{2.5}$  to be energetically more stable than previously investigated structures.  The arrangement and orientation of the TiO$_5$ pentahedra in STO$_{3-\delta}$  is key for tuning the energetic, electronic and magnetic properties. Instead of the previously reported coexistence of localized and delocalized electrons of other STO$_{3-\delta}$ structures, the released electrons are  fully localized   in a semiconducting state with a moment of 1 $\mu_B$/Ti. All released electrons occupy the $d_{xz}$-orbital, leading to an alternating orbital ordering on the  Ti$^{3+}$ sites. The magnetic ordering breaks chiral and time-reversal symmetry and results in a net magnetic moment in the (111) direction. 

This COV phase can explain the experimentally observed ferromagnetism in STO$_{3-\delta}$ as well as many experimental results such as transport properties, structure, and optical absorption/emission. This includes the experimentally observed sixfold superlattice along the (111)$_{\mathrm{cubic}}$ direction with a long space period of 13.52\,\AA~\cite{franco1977anion}. 
The COV phase of STO$_{2.5}$ is expected to form when growing STO along the (111)-direction (due to the hexagonal primitive cell) when synthesized under reduced oxygen pressure or annealed in a vacuum. 
Once synthesized, the chiral arrangement of the $V_{\rm O}$  pentahedra implies that the averaged magnetic moment can only point in the hexagonal out-of-plane or (111)$_{\mathrm{cubic}}$ direction. 
This FM state  can well induce additional side effects when using STO as a substrate. Even if $\delta \ll 0.5$ in the  substrate, it can be  considerably higher in the vicinity of  the interface. This might, e.g., explain the puzzle that magnetic moments exceed $3\,\mu_B$/Ru in   (111)-orientate SrRuO$_3$ films \cite{PhysRevB.85.134429}:  it appears larger because of the additional moment of the COV STO$_{3-\delta}$ substrate.

\section{Acknowledgments}
\begin{acknowledgments}
We thank E.~Maggiao and X.~K.~Ning for helpful comments and discussions.
We acknowledge funding through the Austrian Science Funds (FWF) projects I~5398, P~36213, SFB Q-M\&S (FWF project ID F86), and Research Unit QUAST by the Deuschte Foschungsgemeinschaft (DFG; project ID FOR5249) and FWF (project ID I~5868).
L.~S.~is thankful for the starting funds from Northwest University. M.~K.~was supported by KIAS Individual Grants(CG083501).
Calculations have been done on the Vienna Scientific Cluster (VSC) and super-computing clusters at Northwest University. 
\end{acknowledgments}

\onecolumngrid
\subsection{Supplemental Material to ``Chiral magnetism and ordering of oxygen vacancies in SrTiO$_{2.5}$"}
\onecolumngrid
This supplementary material contains computational details and  additional phonon calculations.
In Section I, we provide computational details and brief descriptions of the methods employed: density-functional theory (DFT) and dynamical mean-field theory (DMFT).
In Section II, we show the phonon spectrum and density of states of the  chiral-oxygen-vacancy (COV) phase of SrTiO$_{2.5}$.
In section III, we show the DFT+$U$ total energies, magnetic orders, magnetic moment and space group of different phases of SrTiO$_{2.5}$.
In section IV, we show the non-collinear magnetic order in the COV phase of SrTiO$_{2.5}$ obtained from DFT+$U$+SOC (spin-orbital coupling) calculations.
In section V, we show the DMFT momentum-resolved spectral function $A(k,\omega)$ of COV phase SrTiO$_{2.5}$.

\section{Section~I: Computational details and analytic continuation} 
\label{sec:details}
DFT structural relaxations and static energy calculations are performed using the \textsc{Vasp} code \cite{PhysRevB.47.558,kresse1996efficiency}. For STO$_3$, the Perdew-Burke-Ernzerhof version for solids of the generalized gradient approximation (GGA-PBESol) \cite{PhysRevLett.100.136406} leads to a better agreement between theoretical (3.907\,\AA) and experimental values (3.905\,\AA) than GGA-PBE (3.949\,\AA) \cite{PhysRevLett.77.3865}. We hence employ GGA-PBESol for DFT(+$U$) calculations. 
The shortcoming of underestimating the band gap of bulk SrTiO$_3$ in standard DFT is solved by employing the modified version of the exchange potential proposed by Becke and Johnson (mBJ) \cite{PhysRevLett.102.226401}.
The Coulomb $U$ for Ti-3$d$ orbital is $U$-$J$=5\,eV (\cite{PhysRevLett.116.157203}) and was employed for the DFT+$U$ approximation using the approach from \cite{PhysRevB.57.1505} in \textsc{VASP} \footnote{Using the approach from Liechtenstein \emph{et al.}~with $U$=5.0\,eV and $J$=0.5\,eV \cite{PhysRevB.52.R5467} does not change the main conclusions in this manuscript}. A dense $k$-mesh of 9$\times$9$\times$9 (11$\times$11$\times$3) with more than 300\,$k$-points is employed for the 2$\times$2$\times$2 supercell (1$\times$1$\times$6 hexagonal supercell) calculations.

As an input for the DFT+DMFT calculations, a low-energy effective Hamiltonian is generated by projecting the Ti-$t_{2g}$ DFT bands computed by \textsc{WIEN2K} around the Fermi level onto Wannier functions \cite{PhysRev.52.191,RevModPhys.84.1419} using \textsc{WIEN2WANNIER} \cite{mostofi2008wannier90,kunevs2010wien2wannier}. These are supplemented by a local density-density interaction with intra-orbital $U$=5\,eV and Hund's exchange $J$=0.5\,eV and the fully localized limit as double counting correction \cite{PhysRevB.44.943}. We solve the resulting many-body Hamiltonian at temperatures from 100\,K - 500\,K employing a continuous-time quantum Monte Carlo solver in the hybridization expansions \cite{RevModPhys.83.349} as implemented in the \textsc{W2dynamics} code \cite{PhysRevB.86.155158,w2dynamics2018}. Real-frequency spectra are obtained with the \textsc{ana\_cont} code \cite{Kaufmann2021}, which performs an analytic continuation through the maximum entropy method (MaxEnt) \cite{PhysRevB.44.6011,PhysRevB.57.10287}.

\section{Section~II: DFT phonons} 
\label{sec:dft_phonons}
\begin{figure*}[h]
\centering
\includegraphics[width=0.8\textwidth]{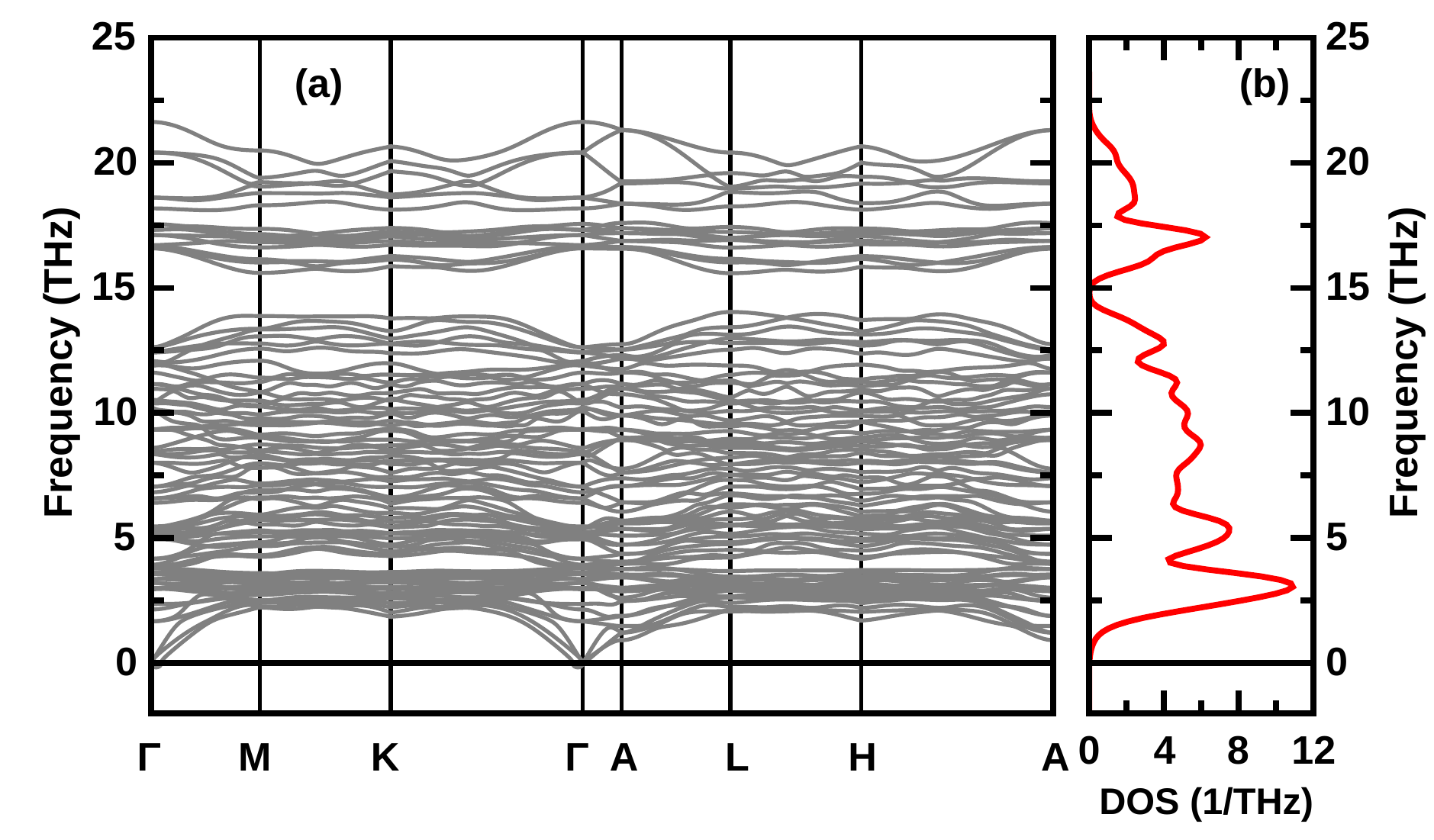}
\caption{Phonon energy-momentum dispersion (a) and DOS (b) for the COV phase SrTiO$_{2.5}$, demonstrating the crystal stability of this phase. Positive phonon frequencies for the whole $q$-path indicate the crystal stability of the COV phase.}
\label{Fig_phonon}
\end{figure*}

Fig.~\ref{Fig_phonon} shows the phonon dispersion of the COV phase STO$_{2.5}$ performed within a 2$\times$2$\times$2 supercell of the COV phase STO$_{2.5}$ [Fig.~1(m) in main text] that contains 216 atoms. For this structure, the frozen phonon method is employed using the \textsc{PHONONY} code \cite{togo2015first} interfaced with \textsc{VASP}. The positive phonon frequencies demonstrate the stability of the COV phase.

\section{Section~III: DFT+$U$ total energy and magnetic moment} 
\label{sec:dft_energy}
The results of GGA-PBESol DFT+$U$ total energy calculations are shown in Fig.~3 of the main text and Table~\ref{Tab1_energy}.
The structures with only TiO$_5$ pentahedra ($A$, $C$, $T$-1, $G$ and $G$-111) have higher energies. The 3$d^1$ electronic configuration of TiO$_5$ pentahedra induces metallic bands crossing E$_f$ for these phases, which are responsible for the energy enhancement. Additionally, in a TiO$_5$ pentahedron crystal field, the single  electron occupies the double degenerate $d_{yz}$+$d_{xz}$ orbitals which leads to a Jahn-Teller distortion, as we discussed in the main text. Other structures in which TiO$_5$ and TiO$_4$ coexists, such as $C$-1, $C$-2, $T$-2 and $T$-3, are energetically between the TiO$_5$ group ($C$, $T$-1, $G$) and TiO$_4$-plus-TiO$_6$ group ($VC$-1, $VC$-2).
Please note that in both DFT and DFT+$U$ calculations, the total energies of $G$ and $G$-111 are identical as they are structurally equivalent.

\begin{table*}[h]
\caption{Non-spin-polarized DFT (Non-mag.) and spin-polarized (DFT+$U$)  total energies, magnetic ground states, space group and magnetic moment (per Ti) from DFT+$U$ for the different STO$_{2.5}$ phases of Fig.~1 in the main text. The total energy of the $VC$-1 phase is set to zero for better comparison. The total energies and magnetic moments (FM: ferromagnet, $A$-, $C$-, $G$-AFM: $A$-, $C$- and $G$-type antiferromagnet) are in units of meV/STO$_{2.5}$ and $\mu_B$/Ti, respectively. The boldface texts indicate the ground state total energies in non-spin-polarized DFT and spin-polaried DFT+$U$ calculations.}
\begin{tabular}{c|c|c|c|c|c|c}
\hline
\hline
Phase    & $VC$-1 & $VC$-2 & $A$ & $C$ & $C$-1 &  $C$-2  \\
Space group    & $P4/mmm$ (123) & $Pmmm$ (47) & $P4/mmm$ (123) & $Cmmm$ (65) &  $P4_{2}/mmc$ (131) & $Pmmm$ (47) \\
\hline
Non-Mag. & $\mathbf {0.00}$ & 0.62 & 37.38 & 185.26 & 84.65 & 109.43 \\
\hline
DFT+$U$  & 0.00 & -43.71 & 146.43 & 211.78 & 99.57 & 102.85    \\
Moment & FM (0.3) & FM (0.4)  &  Non-Mag. (0.0)  & Non-Mag. (0.0)  & $C$-AFM (0.2) & $C$-AFM (0.3)   \\ 
\hline
\hline
Phase    & $T$-1 & $T$-2 & $T$-3 & $G$ &   $G$-(111) & COV  \\
Space group    & $P4/mmm$ (123) & $P4/mmm$ (123) & $Pmmm$ (47) & $I4/mmm$ (139) &   $I4/mmm$ (139) &  $P3_221$ (154)  \\
\hline
Non-Mag. &  178.11 & 77.56 & 52.68 & 290.76 & 290.76 & 211.20  \\
\hline
DFT+$U$   & 217.93 & 108.32 & 59.26 & 337.90 & 337.90 & $ \mathbf {-75.42}$    \\
Moment  & $A$-AFM (0.3)  &  FM (0.2)  & FM (0.3)  & Non-Mag. (0.0) & Non-Mag. (0.0) & FM (1.0)   \\ 
\hline
\hline
\end{tabular}
\label{Tab1_energy}
\end{table*}

\section{Section~IV: Magnetic moment direction of COV phase SrTiO$_{2.5}$} 
\label{sec:magnetism}
\begin{figure*}[t]
\centering
\includegraphics[width=0.5\textwidth]{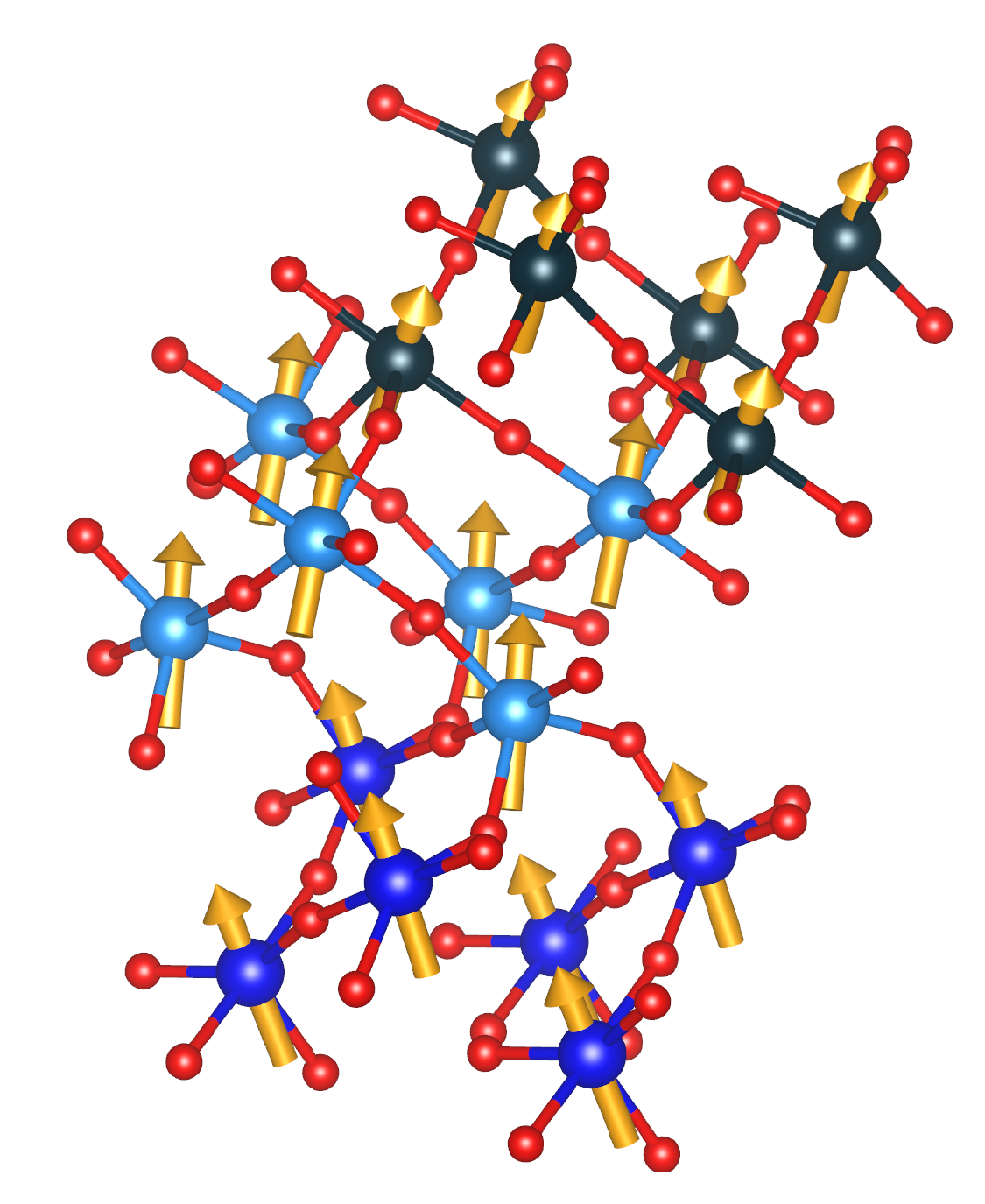}
\caption{Local magnetic moment vectors projected onto Ti cations from FM DFT+$U$+SOC calculations. The Ti cations from 1st and 2nd, 3rd and 4th, 5th and 6th Ti layers are labelled by different blue colors to indicate the Ti-V$_{\mathrm O}$-Ti dimer.}
\label{FigS2_SOC}
\end{figure*}

The directions of magnetic moments on Ti are obtained from DFT+$U$+SOC calculations by setting the moment along different possible directions. The most stable direction is shown in Fig.~\ref{FigS2_SOC}. The converged direction of magnetic moment on Ti are basically parallel to Ti-V$_{\mathrm O}$-Ti direction but host a slight tilting to the out-of-plane direction (111) of cubic coordination.

\section{Section~V: momentum-resolved DMFT spectra of COV phase SrTiO$_{2.5}$} 
\label{sec:Akw}
\begin{figure*}[t]
\centering
\includegraphics[width=0.95\textwidth]{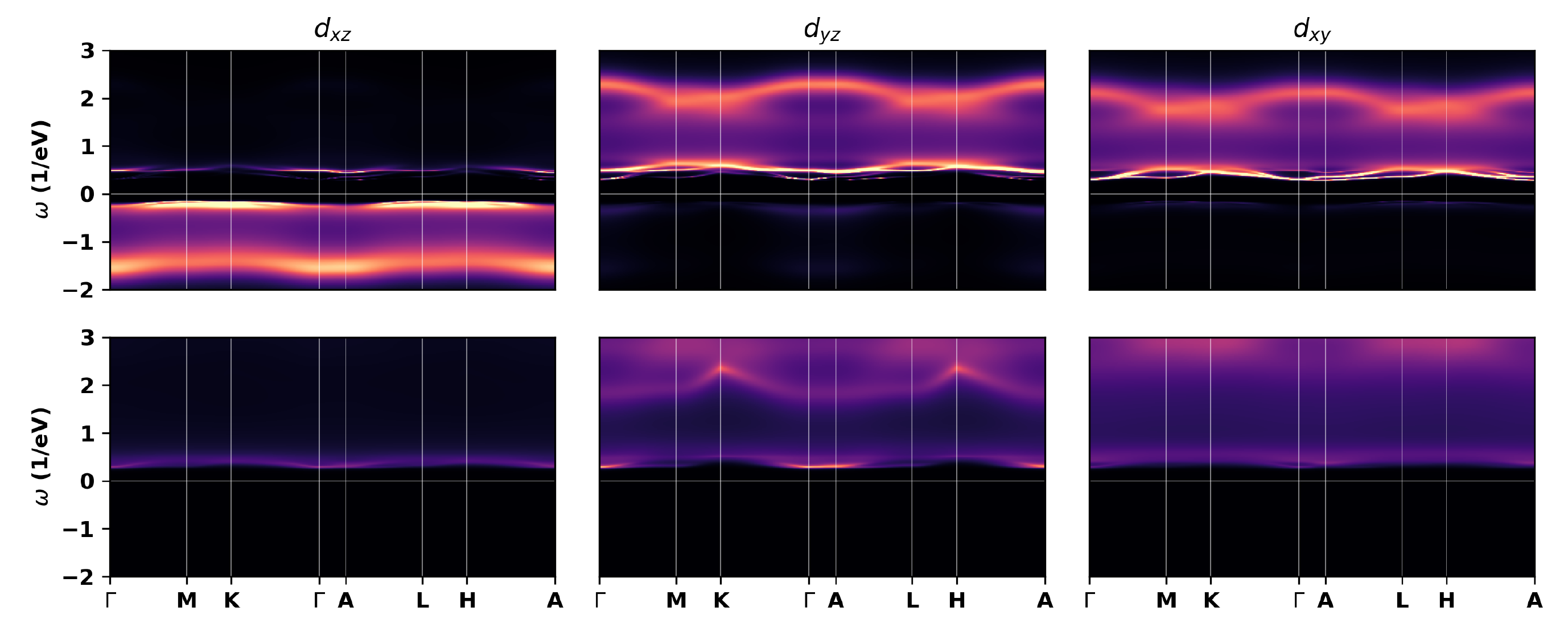}
\caption{Momentum-resolved DMFT spectra $A(w,\omega)$ of COV phase SrTiO$_{2.5}$. The up and down panels indicate spin-majority and minority channels, respectively.}
\label{FigS3_Akw}
\end{figure*}

Fig.~\ref{FigS3_Akw} shows the momentum-resolved DMFT spectra $A(k,\omega)$ of the SrTiO$_{2.5}$ COV phase. We set the spin-up channel as spin-majority. The spin-up (majority) channel of $d_{xz}$ orbital is almost fully occupied, while the $d_{yz}$ and $d_{xy}$ orbitals are empty, leading to a magnetic moment of $\sim$1.0\,$\mu_B$/Ti. The electron excitation (absorption) from the occupied states at -0.3\,eV ($d_{xz}$) and -1.3\,eV ($d_{xz}$) to the unoccupied states at 0.5\,eV ($d_{yz}$ and $d_{xy}$) and $\sim$2\,eV ($d_{yz}$ and $d_{xy}$) are expected to play dominate roles of light emission around 0.8\,eV and 1.8-3.3\,eV. Mostly, the lower Hubbard bands at -1.2\,eV (spin-up of $d_{xz}$) and higher Hubbard bands at 1.8\,eV (spin-up and down of $d_{xy}$ and $d_{yz}$) seem the main contribution to the experimental observation of blue light emission.

\end{document}